\documentstyle[aps,prl]{revtex}
\begin{document}

\twocolumn[\hsize\textwidth\columnwidth\hsize\csname@twocolumnfalse\endcsname

\title{Dipole-quadrupole interactions and the nature of phase III of compressed 
hydrogen}

\author{
        Jorge Kohanoff~$^{1}$, Sandro Scandolo~$^{2,3}$, 
        Stefano de Gironcoli~$^{2,3}$ and Erio Tosatti~$^{1,2,3}$
       }

\address{
         $^{1)}$ International Centre for Theoretical Physics, \\
                 Strada Costiera 11, I-34014 Trieste, Italy \\
         $^{2)}$ International School for Advanced Studies (SISSA) \\
                 via Beirut 4, I-34014 Trieste, Italy\\
         $^{3)}$ Istituto Nazionale Fisica della Materia (INFM)
        }
%\date{\today}
\maketitle

\begin{abstract}
A new class of strongly infrared (IR) active structures is identified for 
phase III of compressed molecular H$_2$ by constant-pressure {\it ab initio} 
molecular dynamics and density-functional perturbation calculations. These 
are planar quadrupolar structures obtained as a distortion of low-pressure 
quadrupolar phases, after they become unstable at about 150 GPa due to a 
zone-boundary soft phonon. The nature of the II-III 
transition and the origin of the IR activity are rationalized by means of 
simple electrostatics, as the onset of a stabilizing dipole-quadrupole 
interaction.
\end{abstract}

\vspace{0.5cm}
]

The nature of the high-pressure phases of solid hydrogen is of long time
interest\cite{wigner}. One of the most intriguing and outstanding issues is
the nature of the 150 GPa phase transition~\cite{phaseIII} and the origin 
of the strong infrared vibronic activity observed in the high pressure 
phase~\cite{hanfland}. The present view of the low temperature phase 
diagram of hydrogen is the following. In the gas and diluted phases the 
homopolar H$_2$ molecules interact via electric quadrupole-quadrupole 
interactions (EQQ). At low temperatures, quantum rotational effects 
overcome librational barriers, so that the 
molecules rotate freely. Upon compression, EQQ interactions freeze
the molecular rotations into an ordered broken-symmetry phase 
(BSP, or phase II). Above $\sim 150$ GPa, a third phase (H-A, or phase III) 
is observed. Optical measurements across the II-III phase boundary indicate 
a substantial increase in the vibron infrared (IR) activity, the  
effective charge being one order of magnitude larger in phase III than in 
phase II~\cite{hemley:epl}. The lack of a Drude edge in extended IR studies 
down to 1000 cm$^{-1}$ indicates that phase III is still insulating~\cite{silvera2}. 
More recent optical experiments focused on the low-frequency 
region of the vibrational spectrum~\cite{goncharov}, revealed the sudden 
appearance in phase III of at least twelve modes below 1000 cm$^{-1}$, 
whose frequencies strongly increase with pressure, extrapolating to zero 
at $P \leq 150$ GPa, and suggesting that quantum librational effects are 
frozen out and are not of paramount importance in phase III~\cite{mazin2}.

A satisfactory theory of the origin of the IR activity and
the nature of the II-III phase transition is still unavailable.
First-principles calculations, although numerous, have failed so 
far to provide a unique description of phase III 
\cite{edwards,ashcroft,klug,alavi,mrs,nagara}, even after inclusion of 
proton quantum effects \cite{surh,natoli,biermann}. Some of them failed in
properly sampling the electronic Brillouin zone~\cite{klug,natoli,biermann},
others in considering unit cells too small to be consistent with the
multiplicity of low frequency 
modes~\cite{edwards,ashcroft,alavi,mrs,nagara,surh}. In particular, some 
structures studied were metallic~\cite{mrs}, and the $Cmc2_1$ structure, 
which is currently being considered as a promising candidate for phase 
III~\cite{ashcroft,nagara,souza}, turns out to be unstable against $Pca2_1$, 
at least within LDA and GGA.

In this letter we report the results of a combined structural and vibrational
study using constant-pressure ab initio molecular dynamics simulations (CPAIMD) 
\cite{focher} and density functional perturbation theory (DFPT) 
calculations \cite{baroni}. These show that, starting with the
likeliest structure proposed for phase II, a zone-boundary phonon 
instability sets in at about 150 GPa. The resulting stable structures
have larger unit cells and exhibit enhanced IR activity, precisely the
features typical of phase III. We then show that the transition from 
low to high IR activity can be understood in terms of simple electrostatics, 
as a structural transition driven by dipole-quadrupole interactions.

The details of the ab initio calculations are the following. 
Electronic correlations were treated within a gradient corrected density 
functional~\cite{beckper}, while the proton-electron interaction was 
described through a local pseudopotential~\cite{giannozzi} that requires 
an energy cutoff of 60 Ry for the plane wave expansion. Convergence on 
BZ sums required 512 $k$-points in the full BZ~\cite{monkhorst} of 4 and 
6-molecule supercells, and 256 $k$-points in an 8-molecule supercell. 
In the MD simulations, full BZ sampling was implemented using a 
self-consistent $k\cdot p$ technique~\cite{mrs,koha,kdp}. 

Previous CPAIMD simulations indicate that the quadrupolar $Pca2_1$, $hcp$ 
structure (see Fig. 1) is likely to correspond to the low-pressure phase 
II~\cite{koha}, confirming that the EQQ are dominant at low densities. In the 
present work we have computed the zone-center and zone-boundary ab initio phonon 
frequencies at 140 GPa using DFPT, and found that this structure is, in 
fact, stable. However, increasing pressure to 180 GPa, the phonon 
dispersion of the $Pca2_1$ structure displays an instability at the (100) 
Brillouin zone boundary~\cite{inconm}. Using the corresponding eigenvector 
we distorted the original structure and minimized the enthalpy by 
optimizing lattice parameters and molecular orientations. In the 8-molecule 
structure derived from the $Pca2_1$, and shown in Fig. 1, we found a 
tendency of the molecules to lie in parallel, (001) $hcp$ planes. This 
tendency becomes more pronounced the higher the pressure.

Once molecules lie in, or close to planes, it is natural to surmise that 
another possible candidate would be a structure that optimizes the 
quadrupolar interactions in a triangular, two-dimensional system. Within 
the planes, neighboring molecules are oriented at 60 degrees from each other. 
This minimizes frustration of EQQ interactions, an angle of 90 degrees being 
impossible in a triangular lattice. Thus, we also optimized the structure 
of the triangular structure by means of CPAIMD, in a 6-molecule cell (see Fig. 1), 
and found it to be stable, with an enthalpy lower than that of the 
distorted $Pca2_1$. In this way we obtained a new class of structures which
are characterized by unit cells containing {\it more} than 4 molecules. 

In order to make a direct comparison with experimental data, we report in Fig. 2
the vibrational spectrum at zone center of the undistorted $Pca2_1$ structure at 
140 GPa and, at higher pressures, where this structure becomes unstable, the 
spectrum of the centrosymmetric triangular structure. We first notice that the 
discontinuous frequency drop upon transformation from the $Pca2_1$ to the triangular 
structure, compares favorably with an experimental jump of about 100 cm$^{-1}$. This 
jump is connected with an intramolecular bond softenining signalled by a sudden, 
$\approx$ 2\% {\it increase} in the molecular bond lengths. Actually, in the triangular 
structure, the bond length is 1.41 \AA for two of the six molecules in the unit cell, 
and 1.39 \AA for the other four. In the $Pca2_1$ structure the bond length is 1.37 \AA. 
Concomitantly, we find a $\approx$ 1\% discontinuous volume decrease at the transition.
In fact, the main energy decrease at the transition is in the interproton repulsion, 
counteracted by a nearly equal increase of electronic energy. The lower part of the 
experimental IR and Raman spectra of phase III shows a number of modes whose 
frequencies extrapolate to zero for decreasing pressure, between 100 and 150 GPa, but 
are totally absent in phase II \cite{goncharov}. Since optics detects only zone-center 
modes, these data strongly suggest a zone-boundary libron instability of phase II, 
with formation of a larger supercell and soft mode refolding at $k=0$ in phase III. 
This is precisely what our first-principles calculations predict. The mode-hardening 
with pressure which we calculate for the lowest frequency mode is 
$41.5\sqrt{(P-142)}$ cm$^{-1}$ (see Fig. 2).

The DFPT approach also allows us to compute IR activities~\cite{effective}, which 
we report in Table I, as a function of pressure. At low pressures three of the 
vibrons are nominally IR active, but the oscillator strengths are very small, with 
effective charges $Z^*\approx 0.02~e$. After the transition to the 
high-pressure phase, the IR activity increases suddenly by one order 
of magnitude, i.e. $Z^*\approx 0.2~e$. This large jump appears to account
quantitatively for the experimental onset of strong IR activity in 
phase III  \cite{hemley:epl}. The IR activity is further shown, using 
effective charges calculated at 220 GPa, to grow steadily by about 
$1.5\times 10^{-4}~e/$GPa with increasing pressure, to be compared with 
an experimental increase of $2.07\times 10^{-4}~e$/GPa. 
Finally, there is an IR phonon mode at about 1600 cm$^{-1}$, 
which may explain that seen in one set of experiments \cite{silvera2}. 

An intriguing fact is that we find three IR active vibrons against only one
observed experimentally. We note here that this threefold multiplicity 
is of strictly interplanar nature. Interplanar interactions cause, e.g. the
splitting between modes A (IR active) and B (Raman active), as indicated in
Fig. 2. If angular correlations between planes were to be reduced, or eliminated,
for example by librational fluctuations, then the AB doublet would collapse,
and the IR activity of mode A would be supressed. The alternative possibility of a
smaller unit cell (3 molecules) is not supported by our calculations, and seems 
also incompatible with the large number of experimental librational modes. 

We computed, in addition, the static dielectric constant, which is experimentally 
accessible via refractive index measurements~\cite{evans}. We predict a discontinuous 
increase of about 10\% at the 150 GPa phase transition, which is at variance with
the decrease calculated for other simpler structures on the basis of a dielectric 
catastrophe mechanism, driven by diverging dipole-dipole interactions \cite{ashcroft}.

We now address the physics underlying the II to III transition, which appears
to be driven by dipole-quadrupole interactions. In fact, a simple electrostatic 
model consisting of interacting quadrupoles and corresponding induced dipoles 
in a $hcp$ lattice qualitatively accounts for the picture emerging from the 
above first-principles results. This model is an extension of the original model of 
interacting pure quadrupoles\cite{miyagi} with the addition that the H$_2$ 
molecules are polarizable, and develop a dipole moment when placed in an external 
field such as that resulting from all the quadrupoles of the neighboring molecules. 
Once these induced dipole moments set in, dipole-quadrupole (EDQ) and 
dipole-dipole (EDD) interactions add to the basic EQQ interactions.
These depend differently on the intermolecular distance ($1/r^5$ for EQQ, 
$1/r^4$ for EDQ, and $1/r^3$ for EDD), thus originating a competition between 
different orientational patterns~\cite{model}. 

In Fig. 3 we show the energy diagram and the behaviour of the dipole 
moment as a function of the lattice constant.
When the molecules are far apart the EQQ interaction is dominant and the 
Miyagi-Nakamura result is recovered \cite{miyagi}: in the $hcp$ 
structure for the molecular centers, a  $Pca2_1$ structure is favored,
the molecules arranging themselves as close to perpendicular as
they can. As the distance between molecular centers decreases,  
all individual quadrupolar electric fields on a given molecule 
increase, inducing a molecular dipole moment. The EDQ interaction, 
is however still weak in $Pca2_1$ because in this 
structure the quadrupolar fields due to neighboring molecules almost 
cancel, and the main contribution to the energy remains EQQ. 
With decreasing spacing, the increasing dipole moment would 
eventually drive a dielectric instability\cite{ashcroft}, as the electric 
fields created by the induced dipoles interact cooperatively. 
As any two molecules approach under pressure, however, the increasing EDQ 
interaction causes their relative orientation angle to turn, driving
a $k\ne 0$ libron instability and eventually stabilizing a new planar 
triangular structure. The dielectric instability is therefore preempted by 
a structural transition. Remarkably, the structure so obtained is very close 
to the planar quadrupolar structure with in-plane quadrupolar orientations 
just surmised from ab initio calculations~\cite{shift}.

The quadrupolar electric field at each molecular centre is now between 5 and 
15 times larger than in the $Pca2_1$ structure due to incomplete cancellation 
of the contributions of the neighboring molecules. This field induces 
proportionally larger dipole moments, so that the EDQ interaction turns out 
to be more than two orders of magnitude larger, and comparable in strength to 
the EQQ interaction. At the transition distance, the static molecular dipole 
moment jumps by a factor of about 8, closely paralleling that of the dynamical 
dipole calculated by DFPT, which is related to the IR activity. The jump of IR 
activity thus signals a very central aspect of the physics driving the II-III 
transition.

In conclusion, we have presented a theory that sheds light on the nature of phase 
III of solid molecular hydrogen, and explains the origin of its strong IR activity. 
The quadrupolar electric field of the H$_2$ molecules induces a dipole moment in 
the {\it polarizable} neighboring molecules. The ensuing dipole-quadrupole 
interaction, gives rise to a nonuniform librational
instability driving a phase transition to a planar structure, still
exhibiting in-plane quadrupolar ordering, but also optimizing the new
term. The corresponding candidate structures obtained from 
first-principles calculations possess larger unit cells, and reproduce 
many measured properties, including the vibronic effective
charges, the vibron discontinuities, the librational phonon spectrum,
and the insulating character of this phase. The exact symmetry group
of phase III and the precise reason for a single IR mode remain open questions,
calling for further theoretical and experimental work.

\begin{figure}
\caption{Schematic view of the low-pressure quadrupolar phase $Pca2_1$ (a), 
and two candidates for phase III: (b) the distorted $Pca2_1$, and (c) the 
triangular structure. Large and small circles indicate that the molecule
lies out of the plane, pointing in the direction of the larger circle. Empty
and shaded molecules correspond to different planes.}
\end{figure}

\begin{figure}
\caption{Zone-center phonon spectrum as a function of pressure for the 
$Pca2_1$ structure at 140 GPa (shifted to the left for the sake of clarity), 
and the triangular structure, above 140 GPa. The pressure dependence of the
IR active vibrons is indicated with dashed lines, and that of the soft 
librational mode with a solid line, crossing zero at $P\approx 142$ GPa.}
\end{figure}

\begin{figure}
\caption{Model of quadrupoles and induced dipoles: Energy (left panel) and 
dipole moment (right panel) per molecule as a function of the lattice constant, 
in an $hcp$ lattice. The dipole moment exhibits a discontinuity of 
nearly one order of magnitude at the transition distance.}
\end{figure}

\begin{table}
\caption{Vibron effective charges as a function of pressure. Values are
reported for the $Pca2_1$ structure at 140 GPa, and for the triangular
structure at 180 and 220 GPa. Experimental values were taken from Ref. [4].}

\begin{tabular}{ccc}
 $P$ (GPa) & Z$^*$ (calc.) & Z$^*$ (exp.) \\ \hline
140 & 0.02 & 0.011 \\
180 & 0.19 & 0.10  \\
220 & 0.21 & 0.13  \\
\end{tabular}
\end{table}


\begin{references}

\bibitem{wigner}  E. Wigner and H. B. Huntington, J. Chem. Phys. {\bf 3},
764 (1935); H. K. Mao and R. J. Hemley, Rev. Mod. Phys. {\bf 66}, 671
(1994) and references therein.

\bibitem{phaseIII} R. J. Hemley, and H.-K. Mao, Phys. Rev. Lett. {\bf 61}, 
857 (1988); H. E. Lorenzana, I. F. Silvera, and K. A. Goettel, Phys. Rev. 
Lett. {\bf 63}, 2080 (1989).

\bibitem{hanfland} M. Hanfland, R. J. Hemley, and H.-K. Mao, Phys. Rev.
Lett. {\bf 70}, 3760 (1993).

\bibitem{hemley:epl} R. J. Hemley, I. I. Mazin, A. F. Goncharov, and
H.-K. Mao, Europhys. Lett. {\bf 37}, 403 (1997).

\bibitem{silvera2} N. H. Chen, E. Sterer, and I. F. Silvera, Phys. Rev.
Lett. {\bf 76}, 1663 (1996); R. J. Hemley {\it et al.}, Phys. Rev. 
Lett. {\bf 76}, 1667 (1996).

\bibitem{goncharov}  A. F. Goncharov, R. J. Hemley, H.-K. Mao, and J. F.
Shu, Phys. Rev. Lett. {\bf 80}, 101 (1998).

\bibitem{mazin2} I. I. Mazin, R. J. Hemley, A. F. Goncharov, M. Hanfland,
 and H.-K. Mao, Phys. Rev. Lett. {\bf 78}, 1066 (1997).

\bibitem{edwards}  B. Edwards, N. W. Ashcroft, and T. Lenosky, Europhys.
Lett. {\bf 34}, 519 (1996).

\bibitem{ashcroft} B. Edwards and N. W. Ashcroft, Nature {\bf 388},
352 (1997).

\bibitem{klug}  J. S. Tse and D. D. Klug, Nature {\bf 378}, 595 (1995).

\bibitem{alavi} A. Alavi, Phil. Trans. R. Soc. Lond. A {\bf 356}, 263 (1998).

\bibitem{mrs} J. Kohanoff and S. Scandolo, Mat. Res. Soc. Proc. Symp. {\bf 499},
329 (1998).

\bibitem{nagara}  K. Nagao and H. Nagara, Phys. Rev. Lett. {\bf 80}, 548 (1998).

\bibitem{surh}  M. P. Surh, T. W. Barbee III, and C. Mailhiot, Phys. Rev.
Lett {\bf 70}, 4090 (1993).

\bibitem{natoli}  V. Natoli, R. M. Martin and D. M. Ceperley, Phys. Rev.
Lett. {\bf 70}, 1952 (1995).

\bibitem{biermann} S. Biermann, D. Hohl, and D. Marx, J. Low Temp. Phys. 
{\bf 110}, 97 (1998).

\bibitem{souza} I. Souza and R. M. Martin, Phys. Rev. Lett. {\bf 81}, 4452 
(1998).

\bibitem{focher}  P. Focher et al., Europhys. Lett. {\bf 36}, 345 (1994).

\bibitem{baroni} S. Baroni, P. Giannozzi and A. Testa, Phys. Rev. Lett. 
{\bf 58}, 1861 (1987).

\bibitem{beckper} The LDA was implemented as in J. P. Perdew and A. Zunger, 
Phys. Rev. B {\bf 23}, 5048 (1981), and gradient corrections as in A. D. Becke, 
Phys. Rev. A {\bf 38}, 3098 (1988); J. P. Perdew and Y. Wang, Phys. Rev. B 
{\bf 33}, 8822 (1986).

\bibitem{giannozzi}  P. Giannozzi (unpublished). See also F. Gygi, Phys. Rev
B {\bf 48}, 11692 (1993).

\bibitem{monkhorst} H. J. Monkhorst and J. D. Pack, Phys. Rev. B {\bf 13}, 5188 
(1976).

\bibitem{koha} J. Kohanoff, S. Scandolo, G. L. Chiarotti and E. Tosatti,
Phys. Rev. Lett. {\bf 78}, 2783 (1997).

\bibitem{kdp}  S. Scandolo and J. Kohanoff (unpublished).

\bibitem{inconm} The minimum of the phonon
dispersion is not necessarily at zone boundary (cell doubling). 
The new stable cell could be larger, or even incommensurate. 
A 3/2 unit cell would correspond to the 
triangular lattice mentioned later in the text.

\bibitem{effective} IR activity is here defined in terms of effective charges,
$Z^*=\mu f\Omega/4\pi$, where $\mu$ is the reduced mass of the molecule,
$f$ the oscillator strength, and $\Omega$ the molecular volume.

\bibitem{evans} W. J. Evans and I. F. Silvera, Phys. Rev. B {\bf 57}, 14105 (1998). 

\bibitem{miyagi} H. Miyagi and T. Nakamura, Prog. Theor. Phys. {\bf 37}, 641 (1967).

\bibitem{model} In order to mimick the H$_2$ molecule, we 
have used a cylindrically symmetric (traceless) quadrupolar tensor with 
a quadrupole moment of 0.5 a.u., and a polarizability tensor with 
a value of 6.7 a.u. for the component parallel
to the molecular axis, and 4.7 a.u. for the orthogonal component 
\cite{kranken}. The model total energy includes EDD, EDQ, EQQ, 
and the electrostatic energy, and is minimized for variable structure and 
spacing. The molecular dipole moments are calculated by assuming a linear
polarizability constitutive relationship.

\bibitem{kranken} J. Van Krankendonk, {\it Solid Hydrogen} (Plenum, NY, 1983),
pp. 15-19. We have taken the $c/a$ ratio of the $hcp$ cell smaller than the 
ideal value, according to former ab initio results \cite{koha}.

\bibitem{shift} The triangular structure obtained from the model differs from
the ab initio one by a sliding of the triangular planes, while the in-plane 
stuctures are identical.

\end{references}
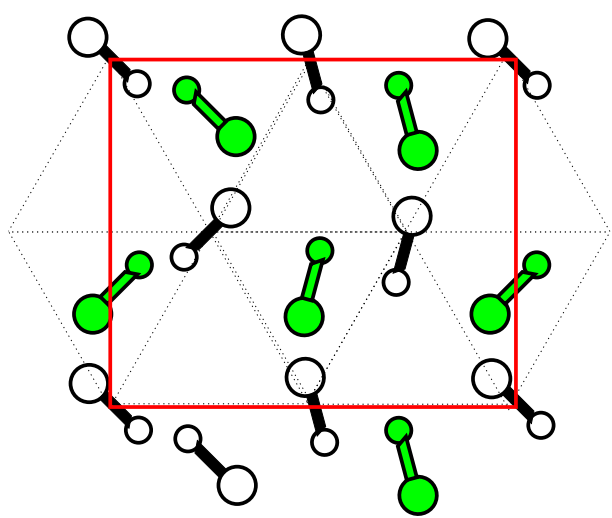
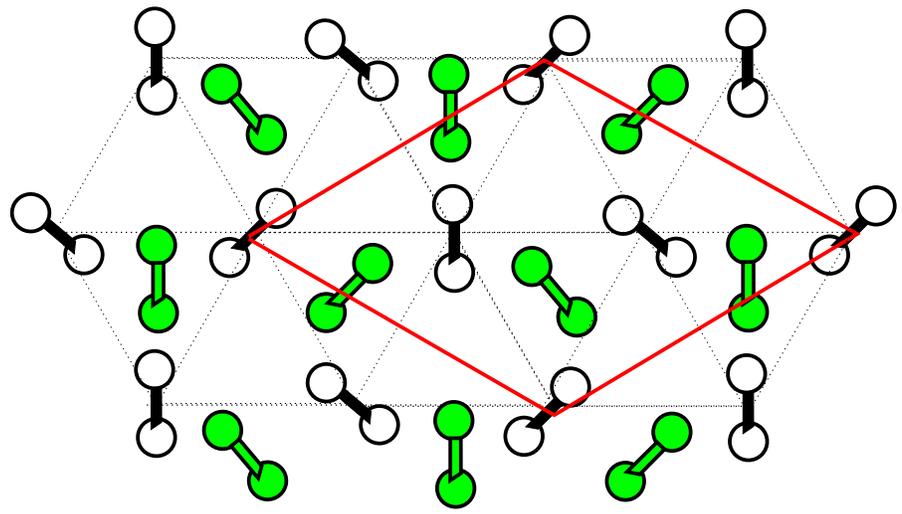
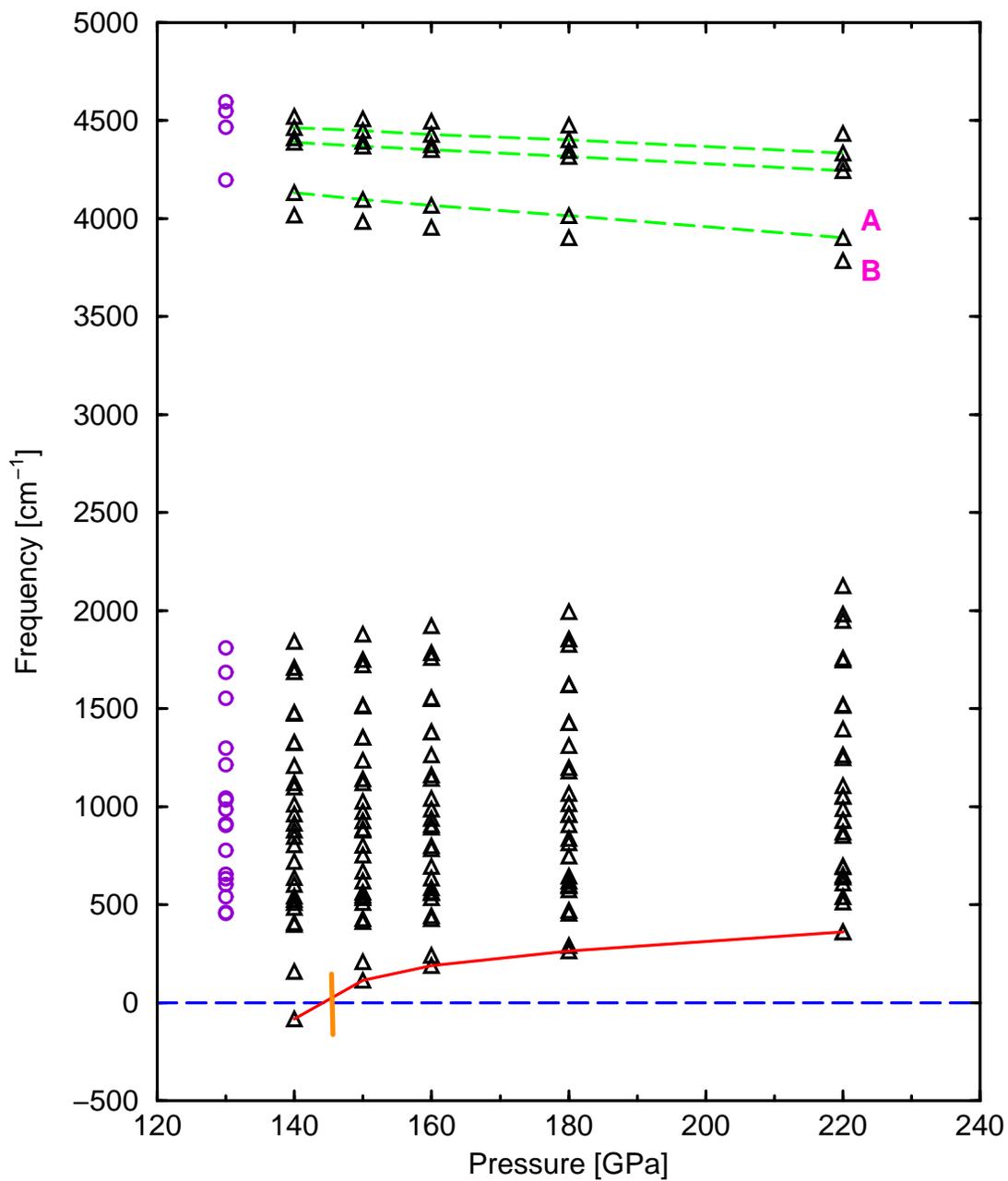
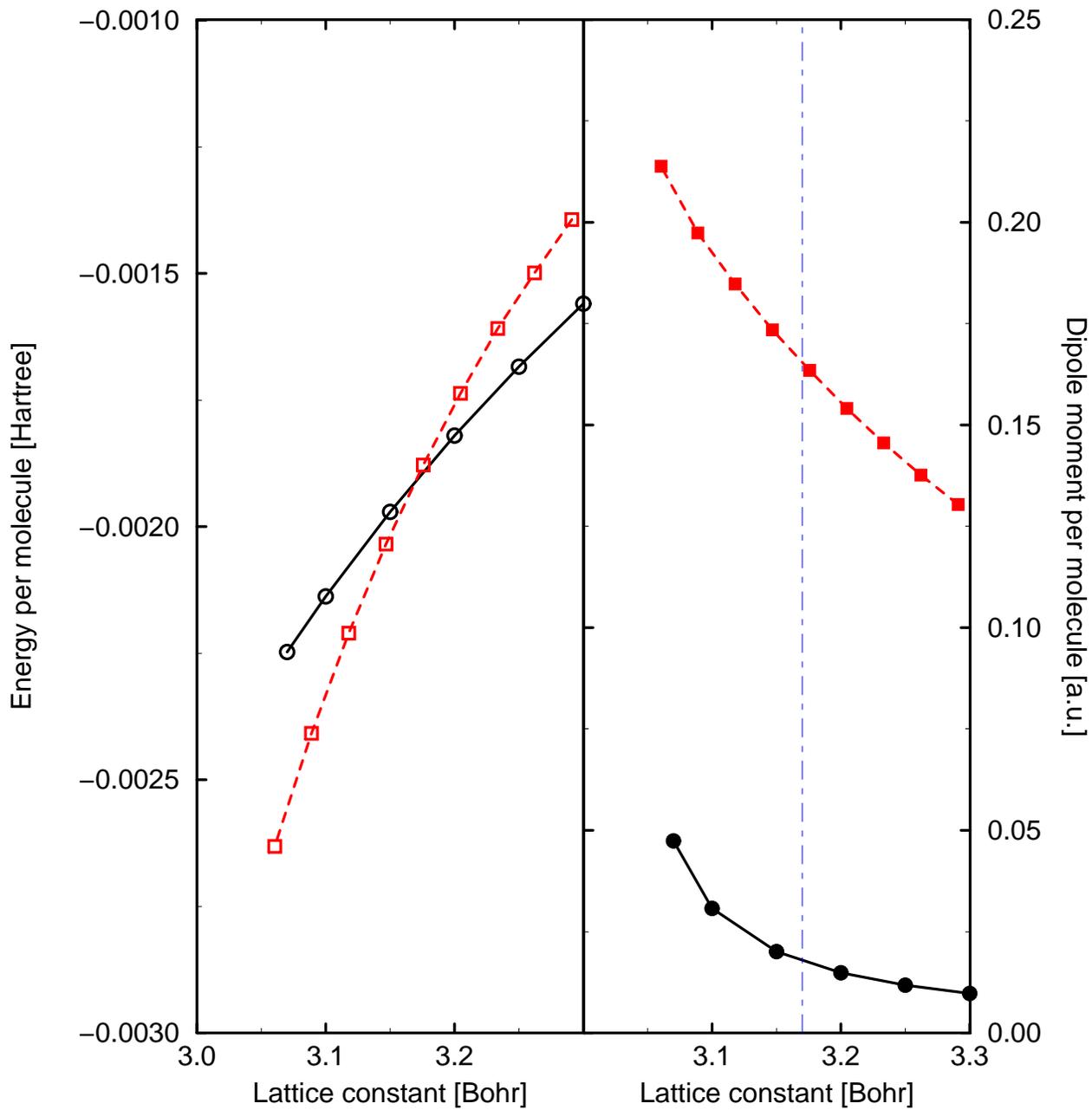
\end{document}